# Coarse-grain Molecular Dynamics Study of Fullerene Transport across a Cell Membrane

Sridhar Akshay, Bharath Srikanth, Amit Kumar[1] and Ashok Kumar Dasmahapatra[2]

Department of Chemical Engineering, Indian Institute of Technology Guwahati, Guwahati – 781 039, Assam, India.

PACS number(s): 87.15.hj, 87.14.Cc, 87.10.Tf, 87.15.nr

---

[1] Corresponding author: Phone: +91-361-258-2275; Fax: +91-361-258-2291; Email address: amitkumar@iitg.ernet.in

[2] Corresponding author: Phone: +91-361-258-2273; Fax: +91-361-258-2291; Email address: akdm@iitg.ernet.in

**ABSTRACT**

The study of the ability of drug molecules to enter cells through the membrane is of vital importance in the field of drug delivery. In cases where the transport of the drug molecules through the membrane is not easily accomplishable, other carrier molecules are used. Spherical fullerene molecules have been postulated as potential carriers of highly hydrophilic drugs across the plasma membrane. Here we report the coarse-grain molecular dynamics study of the translocation of $C_{60}$ fullerene and its derivatives across a cell membrane modeled as a 1, 2-distearoyl-sn-glycero-3-phosphocholine (DSPC) bilayer. Simulation results indicate that pristine fullerene molecules enter the bilayer quickly and reside within it. The addition of polar functionalized groups makes the fullerenes less likely to reside within the bilayer but increases their residence time in bulk water. Addition of polar functional groups to one half of the fullerene surface, in effect creating a Janus particle, offers the most promise in developing fullerene models that can achieve complete translocation through the membrane bilayer.

## I. INTRODUCTION

Every eukaryotic cell is enclosed within a thin membrane called the plasma membrane, which is about 5 nm wide and is the sole pathway for the transfer of substances in and out of the cell[1]. The plasma membrane is predominantly composed of a bimolecular layer of amphiphilic lipids, arranged such that the polar head groups face the surrounding aqueous environment, and the hydrophobic tails face each other[2]. This lipid backbone is infused with proteins and carbohydrates which offer structure and functionality to the membrane[3]. The current accepted model for the membrane structure is the Singer-Nicholson fluid-mosaic model[4]. According to this model, the lipid bilayer is in a fluid state and is free to move laterally within the plane of the membrane. The membrane proteins that penetrate the lipid sheet are located at arbitrary positions, like a mosaic pattern. These components of the membrane are also mobile and are capable of coming into proximity to engage in transient interactions[5].

As the plasma membrane is the sole pathway in and out of the cell, the transfer of drugs and other bioactive molecules across the membrane has attracted significant research interest[6,7]. The direct trans-membrane delivery of drugs is particularly inefficient if the drug molecule is too hydrophilic/polar or if the drug is actively pumped out of the cell. A variety of factors then need to be considered to ensure targeted and effective delivery of drugs to the cells[8].

Drug carrier molecules can generally be classified into four major groups: viral carriers, cationic compounds, proteins and inorganic nanoparticles[9]. Low cell toxicity and ease of availability has made inorganic nanoparticles the focus of much drug research worldwide[10,11]. Fullerene, made up of 60 carbon atoms[12], is one such inorganic nanoparticle that has been investigated for its ability to deliver drugs to their target sites[13–15]. These fullerene molecules cross the bilayer through a combination of passive diffusion and simple hydrophobic forces[16,17]. However, fullerene molecules have remarkably low solubility in water and have shown a tendency to aggregate in polar solutions[18,19]. The pristine fullerene molecules have also been postulated to be cytotoxic[20]. Hence, functionalization of fullerenes is an emerging research area with a view towards improving their solubility[21], reducing cytotoxicity[20] and improving their uptake within the cell membrane[17,22,23].

Molecular dynamics (MD) simulation has emerged as a popular tool for the study and analysis of several membrane-related phenomena[24,25]. Many atomistic MD simulation studies of fullerenes interacting with bilayers have been carried out[26–28]. However, limitations persist with model accuracy due to the extremely slow lateral diffusion of individual lipid molecules within the bilayer. Typical diffusion coefficient for a lipid molecule is around 5 $\mu m^2$/s, which implies that a lipid molecule would take around 50 ns to diffuse from its position to that of an immediately adjacent lipid molecule[29]. Hence, coarse-grain (CG) models are gaining popularity as a method of simulating larger systems over longer timescales. In CG models, small groups of atoms are grouped together as beads and soft-harmonic potentials are used to maintain bond-lengths and angles. This coarse-graining method reduces the number of simulated particles and also allows the usage of larger time-steps. The CG models have been parametrized and used successfully by various groups to study the interaction of carbon nanoparticles and the lipid bilayer[30–32].

Here, we present the results obtained from a coarse-grain MD simulation study on the penetration of $C_{60}$ fullerene molecules and their derivatives into a model 1,2-distearoyl-sn-glycero-3-phosphocholine (DSPC) bilayer membrane. The next section describes the coarse grain models used and the simulation parameters employed in this work.

## II. MODEL AND SIMULATION TECHNIQUE

### A. Coarse-Grain Model

Martini is one of the most popular coarse-grained models for bio-molecular MD simulations[33]. On average, a four-to-one mapping is employed, i.e., four heavy atoms and their associated hydrogen atoms are mapped to a single interaction center. Four main types of interaction centers are defined: Polar (P), Non-Polar (N), Apolar (C) and Charged (Q). Interaction centers P and C are sub-divided into types 1-5, with polarity increasing from 1 to 5. Interaction centers N and Q are sub-divided based on their hydrogen-bonding capability (d - donor, a - acceptor, da - both, 0 - none).

Marrink et al.[34] developed Martini models for the DSPC lipid molecules. The DSPC molecule is made up of an 18 atom long alkyl chain connected to a positively charged ammonium group and negatively charged phosphate group. It is modelled using 14 beads with 10 representing the two lipophilic tails. Two beads represent the oxygen linkages and the charged phosphate and ammonium groups are modelled by one bead each.

The coarse grain model for fullerene used in this work was developed by Monticelli et al.[32,35] by placing 16 beads on the surface of a sphere with a diameter of 0.72 nm and equilibrating using Monte Carlo simulations. The functionalization of fullerene is carried out by replacing the CNP bead defined by Monticelli at al.[32,35] by the more polar P5 particle bead. A range of eight functionalized fullerene models was generated by varying the number and location of the substitutions. Thereby, the dependence of the lipid-fullerene interaction on the degree of polarity and the spatial orientation of the polarity is studied. The various fullerene models used are shown in Figure 1.

Traditional Martini water is defined as a four-to-one mapping with four water molecules represented by a single bead. However, examining the effects of polarity necessitated the use of polarizable Martini water parameterized by Yesylevskyy et al.[36] The parametrization adds two beads to the original water bead and thereby introduces a dipole that interacts more realistically with polar or charged particles.

## B. Simulation Setup

A random orientation of 128 DSPC molecules was solvated in a cubic box of side 7.5 nm. A 30 ns MD run was then used to self-assemble the bilayer. The water molecules used for solvation were removed and the bilayer was oriented in the z-direction. This bilayer-configuration of 128 DSPC molecules was used to represent the membrane in all subsequent simulations.

The fullerenes were then placed at an appropriate distance from the head-groups of the bilayer before being solvated in water. The behavior of the fullerenes in water and their interaction with the bilayer was studied as a function of three different factors - temperature, concentration and spatial polarity. Table 1 provides a description of the fullerene model, number of fullerene molecules and temperature of each simulation run.

A total of 62 simulations were carried out with each run repeated four times. All simulations were carried out for 300 ns with a time step of 30 fs in the NPT ensemble. Temperature and pressure control was maintained through Berendsen thermostat and barostat[37] with constants of 10 ps and 40 ps respectively. The GROMACS simulation software version 4.6.6[38] was used to carry out all the simulations listed in Table 1. Visual Molecular Dynamics (VMD)[39] was used to visualize the results and generate the images.

The energetics of fullerene translocation through the lipid membrane was further studied using a combination of umbrella sampling and weighted-histogram analysis method (WHAM). The initial configuration was generated by placing fullerene at evenly spaced distances of 0.1 nm ranging up to 1 nm from the lipid head-groups in either direction. With a membrane width of 5 nm, this resulted in 70 configurations. The lipid membrane and water molecules were then packed around the configurations. Atomic overlaps were corrected by energy minimization up to the point where the maximum force on any atom was below 10 kJ $mol^{-1}$ $nm^{-1}$. Energy minimization was followed by an NVT equilibration run of 1 ns where the fullerene and the bilayer were restrained by constraints of 2000 kJ $mol^{-1}$ $nm^{-1}$ and 1000 kJ $mol^{-1}$ $nm^{-1}$ respectively. An umbrella potential with a harmonic constant of 1000 kJ $mol^{-1}nm^{-2}$ was then applied to the fullerene and simulated for 10 ns in each configuration. Finally, the GROMACS module **g_wham**[40] was used to perform potential of mean force (PMF) calculation across the windows. The 1D-PMF provides an energetic description of the fullerene as it passes through the membrane.

Two atomistic simulations were also performed to check the reliability of the results obtained by the coarse-grain simulations. Monticelli et al.[32,35] have also developed the atomistic parameters of fullerene for the OPLS all-atom force field[41]. A fully equilibrated 128 molecule 1-palmitoyl,2-oleoyl-sn-glycero-3-phosphocholine (POPC) bilayer developed by Ulmschneider et al.[42] was obtained from Lipidbook[43]. The first atomistic simulation was carried out with only one fullerene molecule placed away from the bilayer and solvated in water. The second atomistic simulation was carried out to verify the fullerene-fullerene interactions and thus had five $C_{60}$ molecules placed outside the bilayer. Both models were simulated for 100 ns in the NVT ensemble with a time-step of 2 fs. The temperature was maintained using the Berendsen thermostat[37] with a constant of 0.1 ps. Figure 2 shows the initial configuration of the fullerene molecules and the POPC bilayer where water molecules have been removed for clarity.

## III. RESULTS AND DISCUSSION

### A. Single Fullerene

Single pristine fullerene molecules tend to spend some time in the bulk of the aqueous phase before diffusing into the bilayer approximately 10 ns into the simulation. The fullerene quickly traverses the head group of the bilayer and moves more slowly through the tail region before settling around the center of the membrane. This behavior of fullerene is qualitatively similar to that observed in simulation studies by D'Rozario et al.[31] and Wong-Ekkabut et al.[32]

#### 1. Effect of Change in Polarity

The functionalization of fullerene causes it to deviate from the ideal behavior described above. This deviation depends on the number and spatial distribution of the polar functional groups on the $C_{60}$ molecule. Figure 3 provides a graph of the variation of the z-coordinate of the fullerene models over time.

Results indicate that the residence time of the fullerene in the bulk solvent is directly proportional to the quantum of polarization. The pristine fullerene molecule resides in the bulk solvent (water) for approximately 10 ns. As the number of polar beads is increased, the residence time in bulk water also increases. At 12 polar beads, the fullerene molecule does not permeate into the bilayer even after 300 ns.

An additional trend is observed in the residence location of the fullerenes after internalization. The pristine fullerene model resides closest to the bilayer center. With increasing polarity, the intra-bilayer residence position of the fullerene models also shifts towards the head-groups. The 12-random fullerene model tends to reside close to the membrane head-groups without permeating. The 12-corner and full-polar models remain completely in the bulk solvent. A notable exception is the Janus particle, which permeates quickly across the membrane and resides close to the head-groups of the lower membrane leaflet. The bulk water residence times and post-internalization residence locations also indicate that functionalized fullerenes with polar group concentrated together interact more strongly with water than fullerenes with more dispersed polarization.

The 1-D PMF values computed using a combination of umbrella sampling and WHAM are shown in Figure 4. It gives the energy of the fullerene models as a function of their position relative to the bilayer center and qualitatively explains the trends observed in Figure 3. The pristine fullerene model shows a deep well with an energy minimum at the center of the bilayer. This is consistent with its quick entry and residence at the bilayer center (as seen in Figure 3). As the quantum of polarization is increased, the depth of the energy well reduces and the minimum point shifts away from the bilayer center, towards the head-groups. This can explain the increased internalization time and residence locations shifting away from the bilayer center in Figure 3. In Figure 3, fullerene models with 12 polar beads or more did not permeate into the bilayer. A similar cutoff is observed in Figure 4 for fullerenes with 12 polar beads or more, where the energy well disappears and the fullerene molecules do not have an 'incentive' to enter the bilayer. However, the PMF calculations are unable to explain the quick diffusion of the Janus particles across the bilayer.

## 2. Effect of Change in Temperature

The dependence of temperature on the behavior of single fullerene molecules was studied using three fullerene models (pristine, polar and Janus). The temperature was varied in intervals of 5ºC from 30ºC to 50ºC. As the temperature of the system is increased, the pristine fullerene molecules exhibit a longer residence time in the bulk-solvent whereas the semi-polar Janus fullerenes show an opposite trend. The average residence time of the two fullerene models is depicted graphically in Figure 5. This temperature dependence can possibly be attributed to the reduction in the dipole moment of water with increasing temperature[44]. However, even increasing the temperature up to 50ºC proved insufficient for the internalization of a fully polar fullerene molecule.

## B. Multiple Fullerene Molecules

When the diffusion of multiple pristine fullerenes is simulated, the $C_{60}$ molecules show a tendency to agglomerate in the aqueous bulk phase before penetrating the bilayer. This fullerene aggregate spends greater time in the aqueous bulk before internalization compared to the single fullerene molecule. This might be due to the smaller surface-to-volume ratio and lower diffusivity of the aggregate.

Post-internalization, the aggregate disperses within the membrane and the fullerene molecules exist individually. Figure 6 depicts the $C_{60}$-$C_{60}$ radial distribution function (RDF) before internalization (at 5 ns) and after internalization (at 40 ns). Pre-internalization, a sharper function is seen with a major peak at $r = 1$ nm. The RDF reaches zero at around 2 nm, which indicates that all the $C_{60}$ molecules are concentrated together. Post-internalization, the RDF is more diffuse and does not reach zero even at large $r$ values, which indicates that the fullerene molecules do not form an aggregated cluster inside the lipid bilayer and are separated from each other. These agglomeration results are qualitatively similar to the simulation behavior observed by Wong Ekkabut et al.[32]

### 1. Effect of Change in Polarity and Concentration

The behavior of functionalized fullerenes is studied using simulation runs of all fullerene models at three concentrations of $C_{60}$ molecules. For pristine fullerene molecules, the increase in concentration causes no deviation in behavior. Even at 15 $C_{60}$ molecules, the fullerenes agglomerate into one or two large particles before transfusing into the bilayer. The larger particles spend a longer time in the bulk water, which can be attributed to the lower diffusivity of the larger mass particles.

As the polarity of the fullerenes is increased, their tendency to agglomerate reduces. Instead of one or two large particles, the functionalized fullerenes form several small aggregates that enter the bilayer at different times. This is possibly due to the competing effects of non-functionalized fullerene beads that promote agglomeration and the hydrophilic polar groups that retard agglomeration. The tendency to form agglomerates reduces with increase in the number of polar beads owing to more favorable interaction with water. Functionalized fullerenes having 12 or more polar beads do not agglomerate at all and reside as individual particles in the bulk water for the duration of the simulation. A similar cut-off in behavior is observed in Figure 3 and Figure 4.

Figure 7 shows the $C_{60}$-$C_{60}$ RDF of the various fullerene models while they reside in the bulk solvent. For pristine fullerene, the peaks in the function occur at low values of '$r$' and the function quickly decays to zero (at $r < 2$ nm). However, as the number of polar functional groups on the fullerene molecules is increased, the RDF progressively becomes broader and decays more slowly to zero, displaying an elongated tail. This behavior of the RDF indicates that although the pristine fullerene particles agglomerate in bulk water, the tendency to agglomerate in water decreases with increase in polarity. The non-polar pristine fullerenes try to minimize the surface exposed to water by agglomerating whereas the polar functionalized fullerenes interact favorably with water and prefer to remain separated instead of forming clusters.

### 2. Effect of Change in Temperature

The effect of temperature on the behavior of fullerene aggregates was studied using three fullerene models namely pristine, Janus and polar. The temperature was again varied in intervals of 5°C from 30°C to 50°C. Simulations with single fullerene molecules showed that increasing the temperature increases the residence time of pristine fullerene in bulk water (see Figure 6). Such behavior would suggest that in simulations with multiple fullerenes, increasing the temperature would reduce the tendency of pristine fullerenes to agglomerate in bulk water. However, no noticeable change in agglomeration behavior was observed with increase in temperature.

The differences in agglomeration and diffusion behavior were within the variations observed between repeated simulations at the same temperature. Even at a temperature of 50°C, the pristine fullerene molecules formed large aggregates, the Janus models to a lesser extent and the polar models did not form aggregates at all. This would suggest that decrease in the dipole moment of water with temperature is not large enough to have any significant effect on the interactions between fullerene molecules. Change in agglomeration behavior might occur at temperatures above 50°C. However, such behavior is difficult to predict, as the exact variation in dipole moment of the MARTINI water model with temperature has not been quantified.

## C. Atomistic Simulations

The atomistic simulations using the OPLS all-atom force field[41] were carried out to validate the results of the coarse-grain simulations. In the atomistic simulations, the residence time of the fullerenes in the bulk water is larger than that observed in coarse grain simulations. In multiple fullerene simulations, the fullerene molecules take a longer time to agglomerate and transfuse into the bilayer. This speeding up of dynamic properties is well known and has been observed extensively in other multi-scale simulations[45,46]. This has been attributed to the softer coarse-grained potentials reducing local friction and energy barriers, thereby accelerating diffusion[47]. However, the trends and behaviors observed in the coarse-grained simulations were qualitatively reproducible in the atomistic runs. Hence, the coarse-grained MARTINI parametrization of DSPC[34], fullerene[35] and water[36] can be considered reasonably accurate for the purpose of this study.

## D. Janus-Particle Diffusion

The translocation of fullerenes from bulk water into the membrane was studied as a function of fullerene models, temperature and fullerene concentration. For fullerenes to be effective as drug carriers, they must be able to pass through the membrane. However, the complete diffusion of fullerenes through the membrane was not observed in any model tested. The fullerene model that showed the most promise was the Janus particle. In the Janus particle, one face of the sphere is polar with particle type P5 while the other face is non-polar with particle type C1. This Janus particle diffuses from the bulk water into the membrane and rests at the tip of the bilayer. The polar hemisphere faces the lipid head-groups while the non-polar hemisphere faces the lipid tails. Figure 8 shows the z-coordinate of the fullerene molecule with membrane head-groups represented by dotted black lines. The total width of the bilayer is around 5 nm and the fullerene molecule resides at a distance of 2.4 nm from the bilayer center.

This orientation behavior is similar to that observed by Ding et al.[48] and Gao et al.[49] in their experimental studies on Janus particles. The quick diffusion across the membrane and the residence close to the bilayer head-group makes it the most promising model for development as a drug carrier. The behavior of Janus fullerenes after their agglomerates migrate inside the membrane is also interesting. Figure 9 provides a snapshot of the fullerenes post-internalization. The agglomerate does not break up as pristine models do, but transforms into a planar form while residing close to the head-groups.

## IV. CONCLUSIONS

The readiness of the fullerene molecules to leave the aqueous bulk phase and enter the lipid bilayer suggests that they can be used as drug carriers. However, the hydrophobic nature of the fullerene molecules makes their escape from within the bilayer energetically unfavorable. The 1D- PMF calculations in Figure 4 and similar calculations by Bedrov et al.[27] and D'Rozario et al.[31] offer weight to this argument. Functionalizing the surface of the fullerene with polar groups can reduce the strength of lipid tail - $C_{60}$ interactions. However, polar functionalization can also have the opposite effect of not allowing fullerene to enter the membrane in the first place. Janus particles[50] present the most promising solution to balancing these opposite forces and achieving complete migration. Further work on Janus fullerenes can involve conducting atomistic simulations to tune the exact strength and spatial orientation of polar surface functional groups in order to achieve complete migration.

Merely increasing the concentration of fullerene in the extracellular domain has been found to be insufficient to cause complete fullerene migration. A larger concentration of extracellular fullerene might cause complete migration; however, this hypothesis was not tested above 15 $C_{60}$ molecules. Work can be aimed at calculating the extracellular concentration large enough to ensure complete migration by passive diffusion. Further efforts can also be directed at a comparative study on the effectiveness of fullerene vis-à-vis other drug carriers such as cell penetrating peptides[51,52]. Studies can also include fullerene models conjugated with the associated drug molecules.

## REFERENCES

[1] G. Karp, *Cell and Molecular Biology : Concepts and Experiments*, 6th Edt. (John Wiley & Sons, 2009), pp. 117–119.

[2] O. Mouritsen and K. Jørgensen, Pharm. Res. **15**, 1507 (1998).

[3] M. Bretscher, Science **181**, 622 (1973).

[4] S.J. Singer and G.L. Nicolson, Science **175**, 720 (1972).

[5] A. Kusumi and Y. Sako, Curr. Opin. Cell Biol. **8**, 566 (1996).

[6] R. Langer and N. Peppas, AIChE J. **49**, 2990 (2003).

[7] G.P.H. Dietz and M. Bähr, Mol. Cell. Neurosci. **27**, 85 (2004).

[8] P. Pàmies and A. Stoddart, Nat. Mater. **12**, 957 (2013).

[9] Z.P. Xu, Q.H. Zeng, G.Q. Lu, and A.B. Yu, Chem. Eng. Sci. **61**, 1027 (2006).

[10] L. Bauer, N. Birenbaum, and G. Meyer, J. Mater. Chem. **14**, 517 (2004).

[11] C. Barbe, J. Bartlett, L. Kong, K. Finnie, H.Q. Lin, M. Larkin, S. Calleja, A. Bush, and G. Calleja, Adv. Mater. **16**, 1959 (2004).

[12] H. Kroto, A. Allaf, and S. Balm, Chem. Rev. **318**, 162 (1991).

[13] J. Shi, H. Zhang, L. Wang, L. Li, H. Wang, Z. Wang, Z. Li, C. Chen, L. Hou, C. Zhang, and Z. Zhang, Biomaterials **34**, 251 (2013).

[14] M.J. Al-Anber, a. H. Al-Mowali, and a. M. Ali, Acta Phys. Pol. A **126**, 845 (2014).

[15] I. Blazkova, H. Viet Nguyen, M. Kominkova, R. Konecna, D. Chudobova, L. Krejcova, P. Kopel, D. Hynek, O. Zitka, M. Beklova, V. Adam, and R. Kizek, Electrophoresis **35**, 1040 (2014).

[16] D. Cai, J. Mataraza, Z. Qin, and Z. Huang, Nat. Methods **2**, 449 (2005).

[17] D. Pantarotto, R. Singh, D. McCarthy, M. Erhardt, J.-P. Briand, M. Prato, K. Kostarelos, and A. Bianco, Angew. Chem. Int. Ed. Engl. **43**, 5242 (2004).

[18] R.S. Ruoff, D.S. Tse, R. Malhotra, and D.C. Lorents, J. Phys. Chem. **97**, 3379 (1993).

[19] N. Sivaraman and R. Dhamodaran, J. Org. Chem. **57**, 6077 (1992).

[20] C. Sayes, J. Fortner, W. Guo, and D. Lyon, Nano Lett. **4**, 1881 (2004).

[21] M. Brettreich and A. Hirsch, Tetrahedron Lett. **39**, 2731 (1998).

[22] R. Partha, L. Mitchell, J. Lyon, P. Joshi, and J. Conyers, ACS Nano **2**, 1950 (2008).

[23] L. Dugan and D. Turetsky, Proc. Natl. Acad. Sci. U. S. A. **94**, 9434 (1997).

[24] P. Coppock and J. Kindt, Langmuir **25**, 352 (2009).

[25] A. Polley, S. Vemparala, and M. Rao, J. Phys. Chem. B **116**, 13403 (2012).

[26] L. Li and H. Davande, J. Phys. Chem. B **111**, 4067 (2007).

[27] D. Bedrov and G. Smith, J. Phys. Chem. B **112**, 2078 (2008).

[28] R. Qiao, A. Roberts, A. Mount, S. Klaine, and P. Ke, Nano Lett. **7**, 614 (2007).

[29] M.J. Stevens, J. Chem. Phys. **121**, 11942 (2004).

[30] X. Shi, Y. Kong, and H. Gao, Acta Mech. Sin. **24**, 161 (2008).

[31] R.S.G. D'Rozario, C.L. Wee, E.J. Wallace, and M.S.P. Sansom, Nanotechnology **20**, 115102 (2009).

[32] J. Wong-Ekkabut, S. Baoukina, W. Triampo, I.-M. Tang, D.P. Tieleman, and L. Monticelli, Nat. Nanotechnol. **3**, 363 (2008).

[33] A.H. De Vries, S.J. Marrink, J. Risselada, D.P. Tieleman, and S. Yefimov, J. Phys. Chem. B **111**, 7812 (2007).

[34] S.J. Marrink, A.H. de Vries, and A.E. Mark, J. Phys. Chem. B **108**, 750 (2004).

[35] L. Monticelli, J. Chem. Theory Comput. **8**, 1370 (2012).

[36] S.O. Yesylevskyy, L. V. Schäfer, D. Sengupta, and S.J. Marrink, PLoS Comput. Biol. **6**, 1 (2010).

[37] H.J.C. Berendsen, J.P.M. Postma, W.F. van Gunsteren, a. DiNola, and J.R. Haak, J. Chem. Phys. **81**, 3684 (1984).

[38] D. Van Der Spoel, E. Lindahl, B. Hess, G. Groenhof, A.E. Mark, and H.J.C. Berendsen, J. Comput. Chem. **26**, 1701 (2005).

[39] W. Humphrey, A. Dalke, and K. Schulten, J. Mol. Graph. **14**, 33 (1996).

[40] J. Hub, B. de Groot, and D. Van Der Spoel, J. Chem. Theory Comput. **6**, 3713 (2010).

[41] W. Jorgensen, J. Am. Chem. Soc. **118**, 11225 (1996).

[42] J. Ulmschneider and M. Ulmschneider, J. Chem. Theory Comput. **108**, 16264 (2009).

[43] J. Domański, P.J. Stansfeld, M.S.P. Sansom, and O. Beckstein, J. Membr. Biol. **236**, 255 (2010).

[44] D. Kang, J. Dai, and J. Yuan, J. Chem. Phys. **135**, (2011).

[45] K.R. Prasitnok and M. Wilson, Phys. Chem. Chem. Phys. **15**, 17093 (2013).

[46] P.K. Depa and J.K. Maranas, J. Chem. Phys. **126**, 054903 (2007).

[47] P. Depa, C. Chen, and J.K. Maranas, J. Chem. Phys. **134**, 014903 (2011).

[48] H. Ding and Y. Ma, Nanoscale **4**, 1116 (2012).

[49] Y. Gao and Y. Yu, J. Am. Chem. Soc. **135**, 19091 (2013).

[50] S. Granick, S. Jiang, and Q. Chen, Phys. Today **62**, 68 (2009).

[51] M. Mazel, P. Clair, and C. Rousselle, Anticancer. Drugs **12**, 107 (2001).

[52] R. Rennert, I. Neundorf, and A.G. Beck-Sickinger, in *Methods Mol. Biol.* (2009), pp. 389–403.

**TABLES**

Table I: Details of the unrestrained MD simulations performed

| No. of Fullerene molecules | Fullerene Models | Temperature (°C) |
|---|---|---|
| 1 | Full-polar | 30;35;40;45;50 |
| 1 | Full-nonpolar | 30;35;40;45;50 |
| 1 | Janus | 30;35;40;45;50 |
| 5 | Full-polar | 30;35;40;45;50 |
| 5 | Full-nonpolar | 30;35;40;45;50 |
| 5 | Janus | 30;35;40;45;50 |
| 5 | All | 30 |
| 10 | All | 30 |
| 15 | All | 30 |
| 1 | All | 30 |

**FIGURES**

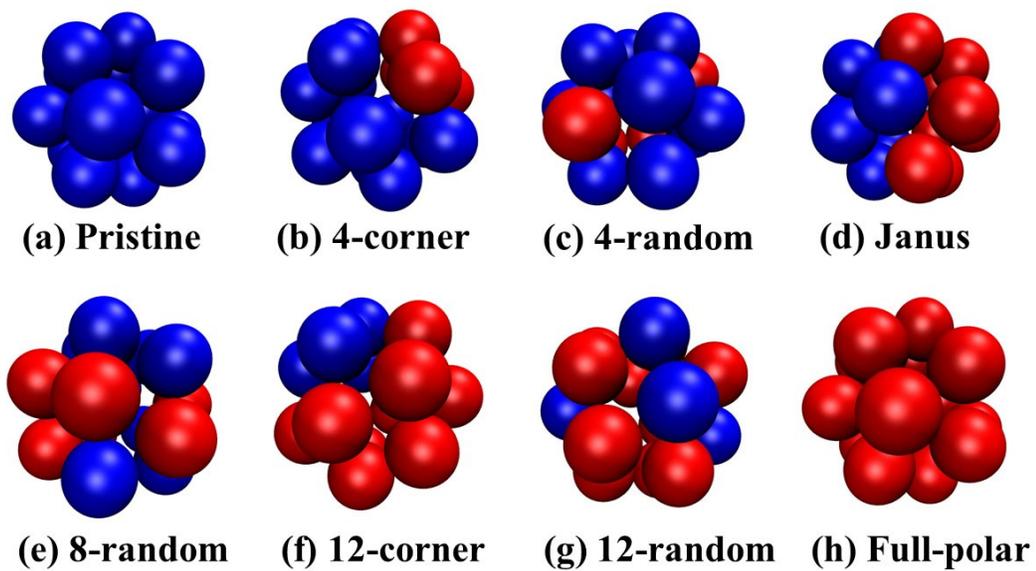

**Figure 1:** Various functionalized fullerene models with non-polar beads in blue and polar beads in red

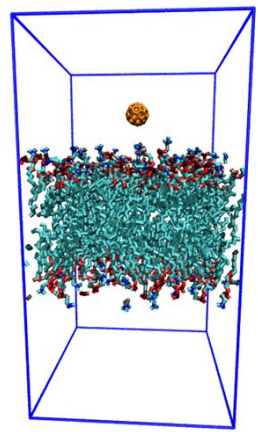 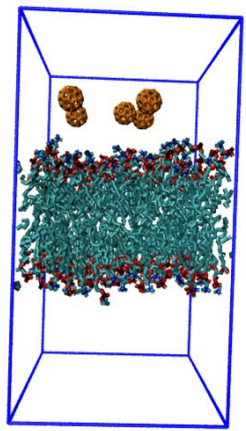

**(a) Single atomistic fullerene with bilayer**

**(b) Five atomistic fullerenes with bilayer**

**Figure 2:** Snapshots of the initial configurations of the membrane with 1 fullerene (left) and 5 fullerenes (right)

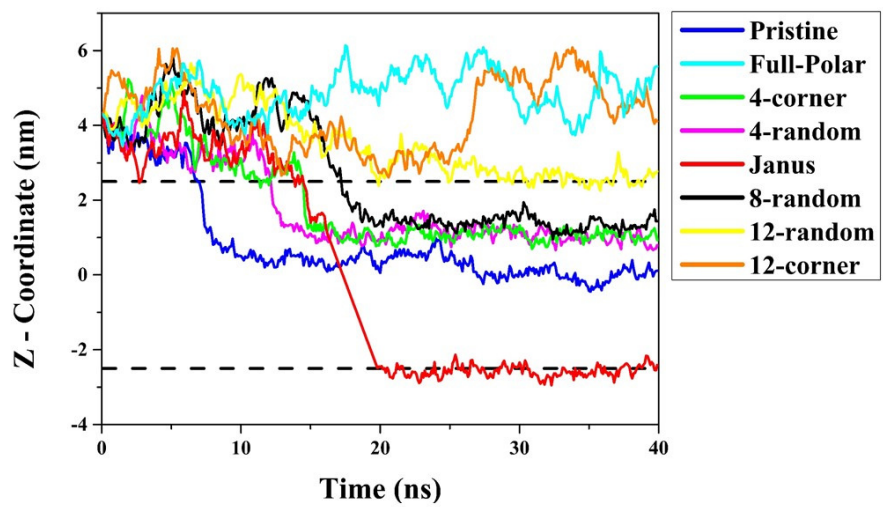

**Figure 3:** Graph detailing the average z-coordinate of the fullerene models as a function of time. The bilayer head-groups are shown as black dotted lines.

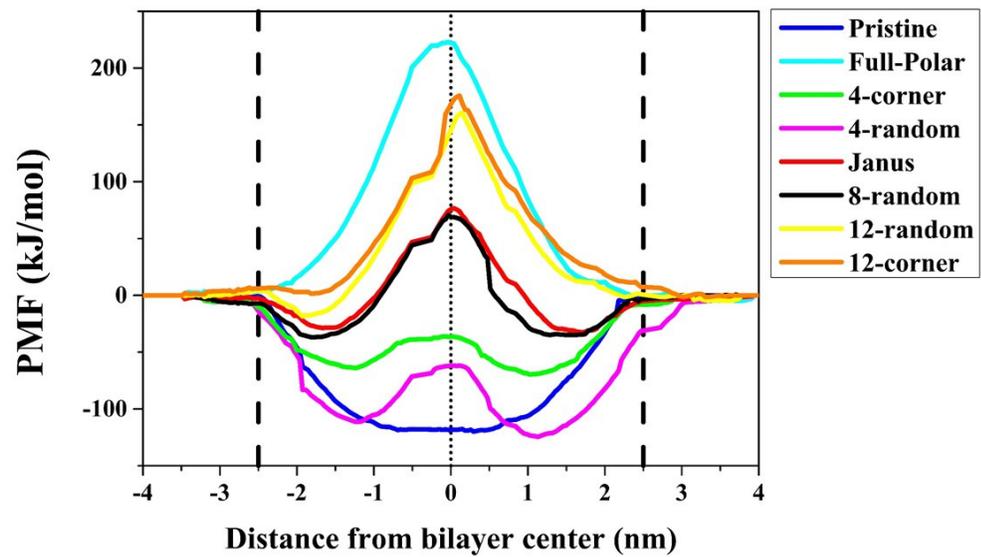

**Figure 4:** Graph showing the 1-dimensional PMF of the fullerene models as a function of z-coordinate across the membrane. Dashed vertical lines indicate the head-groups of the membrane. The dotted line indicates the bilayer center.

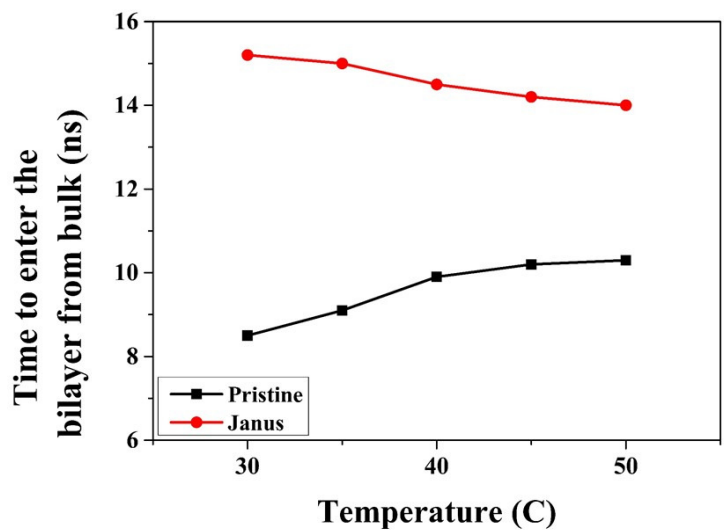

**Figure 5:** Average residence time of single fullerene molecules in bulk water as a function of temperature

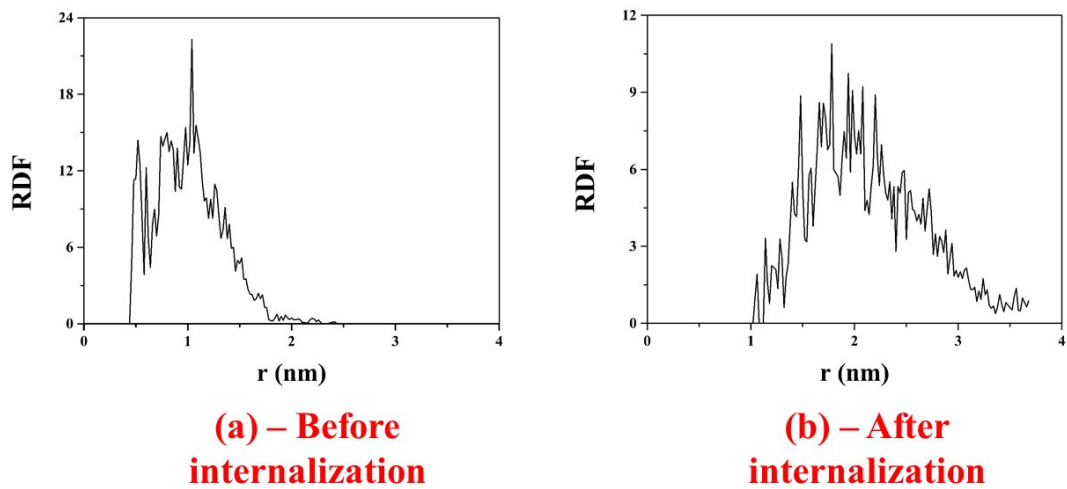

**(a) – Before internalization**

**(b) – After internalization**

**Figure 6:** $C_{60}$ - $C_{60}$ RDF graphs of pristine fullerenes at 5 ns (left) and at 40 ns (right)

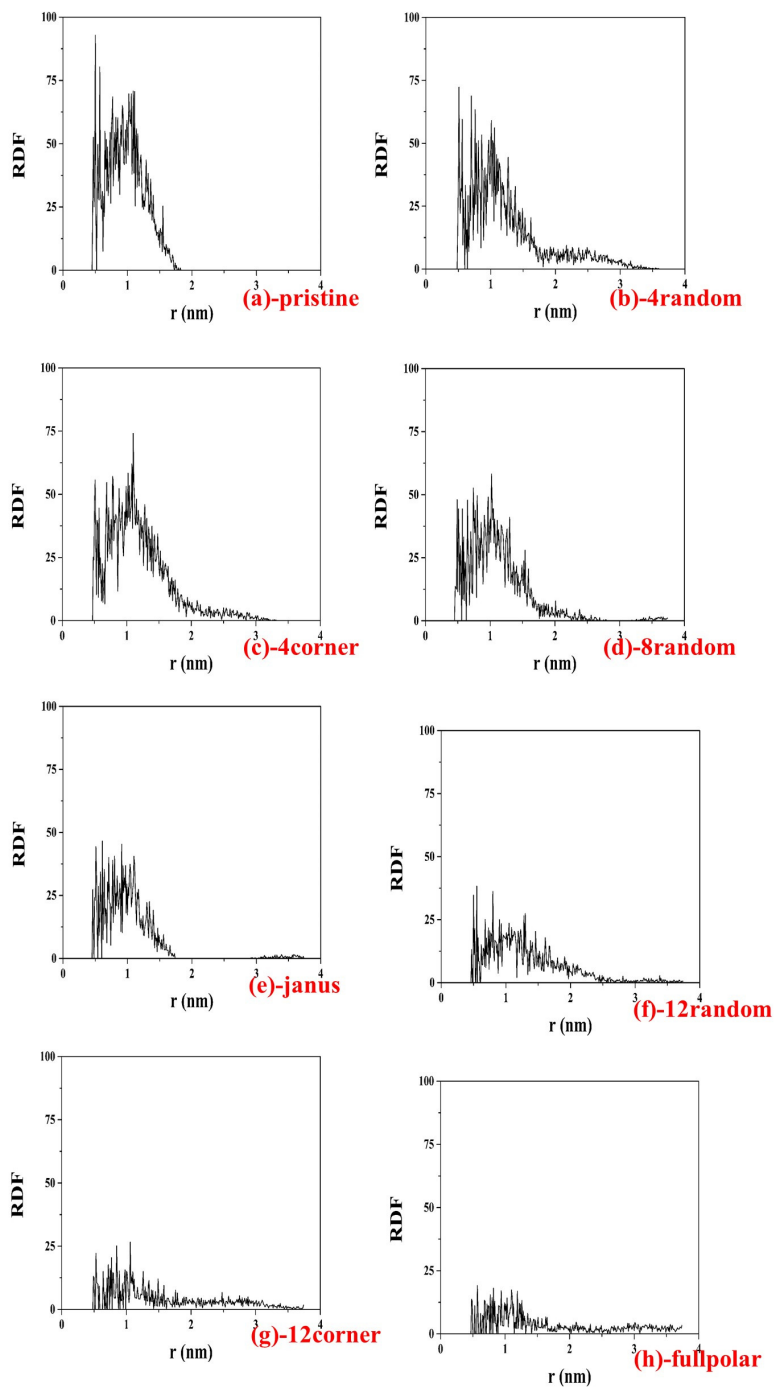

**Figure 7:** $C_{60}$ - $C_{60}$ RDF graphs of fullerene models in bulk solvent. (a) Pristine, (b) 4random, (c) 4corner, (d) 8random, (e) Jauns, (f) 12random, (g) 12corner, (h) fullpolar

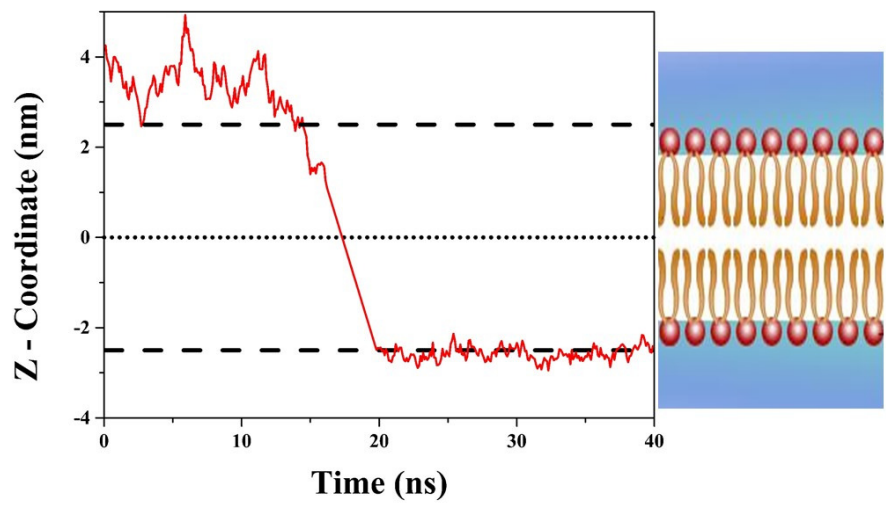

**Figure 8:** Graph showing the z-coordinate of Janus fullerene as a function of time with the membrane head-groups shown as black dotted lines.

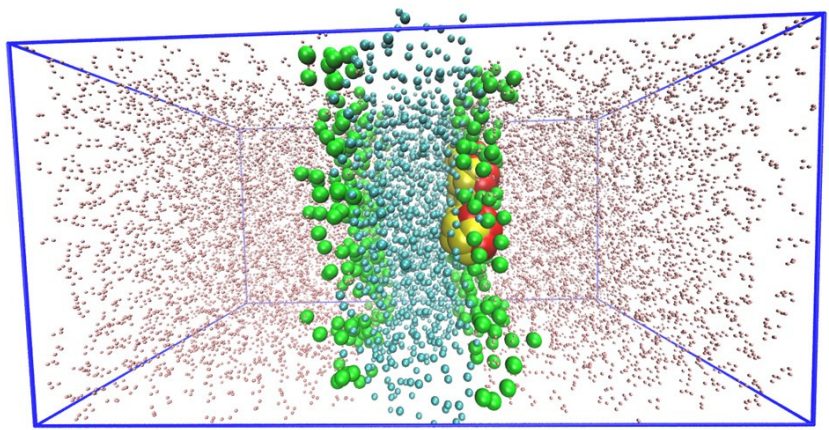

**Figure 9:** Snapshot of Janus fullerenes post-internalization. The lipid tails are shown in cyan and head-groups in green. The polar $C_{60}$ beads are shown in red and non-polar in yellow. Water beads are scaled down and shown as pink specks for clarity.